\documentclass[twocolumn]{cepart}    

\usepackage{amsmath}
\usepackage{amsfonts}
\usepackage{todonotes}
\usepackage{cuted}
\usepackage{etoolbox}
\AfterEndEnvironment{strip}{\leavevmode}
\usepackage{graphicx}          

\usepackage{epstopdf}
\usepackage[round]{natbib}
\usepackage{pgf,tikz,pgfplots}
\pgfplotsset{compat=1.12}
\usepackage{mathrsfs}
\usetikzlibrary{arrows}
\usetikzlibrary[patterns]

\newtheorem{assumption}{Assumption}
\newtheorem{proposition}{Proposition}
\theoremstyle{definition}

\newtheorem{remark}{Remark}

\begin{document}

\begin{frontmatter}

\title{Experimental evaluation of sensor attacks and defense mechanisms in feedback systems}


\author[Stockholm]{David Umsonst}\ead{umsonst@kth.se},    
\author[Stockholm]{Henrik Sandberg}\ead{hsan@kth.se},               

\address[Stockholm]{Division of Decision and Control Systems, School of Computer Science and Electrical Engineering, KTH Royal Institute of Technology, Stockholm, Sweden}  


\begin{abstract}                          
In this work, we evaluate theoretical results on the feasibility of, the worst-case impact of, and defense mechanisms against a stealthy sensor attack in an experimental setup.
We demonstrate that for a controller with stable dynamics the stealthy sensor attack is possible to conduct and the theoretical worst-case impact is close to the achieved practical one.
However, although the attack should theoretically be possible when the controller has integral action, we show that the integral action slows the attacker down and the attacker is not able to remain stealthy if it has not perfect knowledge of the controller state.
In addition to that, we investigate the effect of different anomaly detectors on the attack impact and conclude that the impact under detectors with internal dynamics is smaller.
Finally, we use noise injection into the controller dynamics to unveil the otherwise stealthy attacks.
\end{abstract}

\end{frontmatter}

\section{Introduction}
\label{sec:Introduction}
As more critical infrastructures and industrial processes are connected via communication networks the threat of cyber-attacks on these systems increases rapidly \citep{ListOfCyberattacks}. 
The attacks can range from sophisticated attacks, such as the Stuxnet attack on an Iranian uranium enrichment facility \citep{StuxnetAttack}, to attacks where the attacker takes over the whole control system as in the attack on the Ukranian power grid \citep{UkraineAttack}.
Since these attacks target critical infrastructures, their impact on society can be severe and new methodologies to secure control systems need to be considered \citep{ChallengesForSecurityOfControlSystems}.

As a response to the threat of attacks on control systems, a new field of security complementary to the IT security measures has emerged, which uses physical process knowledge and control-theoretic results to investigate both attack strategies as well as defense mechanisms \citep{TutorialOnSecurityAnnualReview}.
While many attack strategies have been investigated, we focus on sensor attacks in this work, because accurate sensor measurements are important for feedback control. 
Therefore, we want to investigate how a manipulation of these measurements can influence the closed-loop performance.
There exists a large body of work on sensor attacks, see, for example, \citep{FalseDataInjectionMo}, \citep{RuthsMultivariate}, and \citep{LinearSensorAttack}. 
All of these consider a strong attacker model with full model knowledge of the closed-loop system, and the ability to manipulate sensor measurements. 
Furthermore, if a detector is monitoring the system, the attacker also wants to remain stealthy, that is, not trigger an alarm in the detector.
In addition to that, the attacker has also knowledge to internal states of the closed-loop system, such as the controller and the detector state.
However, the knowledge about the internal states is unrealistic in the beginning of the attack. 
In \citep{UmsonstAutomatica21}, we prove that the attacker is only able to estimate the controller state perfectly if and only if the linear controller dynamics do not have any eigenvalues outside the unit circle.
Once the attacker knows the controller state, to inject its stealthy worst-case attack, the attacker needs to determine the detector's internal state if a detector with internal dynamics is used. 
We investigate the problem of detector state estimation in \citep{UmsonstACC19} and conclude that the attacker can estimate the detector state as well, under the assumption of linear detector dynamics.
With knowledge about both the controller and the detector state, the attacker can now inject its worst-case attack. 
Here, we will use the convex worst-case impact estimation problem proposed in \citep{UmsonstACC17} to estimate the worst-case impact of a stealthy sensor attack.
Furthermore, the results on sensor attacks are often theoretical in nature, and might not hold when real data is used.
The need for evaluation of the attack and detection schemes is also pointed in the survey on cyber-physical security by \citet{AttackSurvey}. 
Therefore, the goal of this work is to inspect the impact of the sensor attack in an experimental setup. 
The real process we use in the experiment is the Temperature Control Lab \citep{TCLabModelBenchmark}, which is an Arduino-based process used in (remote) education of control students.
In this work, we focus on the impact on, and defense mechanisms for, the physical system in the loop and not the cyber-aspects of the attack.

The contribution of this work is twofold.
First, we combine three previously independent results, derived in \citep{UmsonstAutomatica21}, \citep{UmsonstACC19}, and \citep{UmsonstACC17}, respectively, into one complete sensor attack strategy with three distinct stages. 
This illustrates that even a powerful attacker with perfect model knowledge and access to the sensor measurements cannot launch a stealthy attack immediately in a realistic scenario, but needs to complete two more stages before it can launch the worst-case attack.
Second, we evaluate the theoretical results of the three-staged sensor attack strategy in an experimental setup with a real process.
We show that if the attack does not complete the first stage successfully, the attack is detected in one of the later stages. 
Further, delaying the completion of the first stage of the attack can be achieved by adding an integral part to the controller. 
The controllers we investigate are the linear quadratic Gaussian (LQG) controller and the LQG controller with integral action.
Moreover, we evaluate how close the theoretical worst-case impact estimation is to the actual impact of the attack and simultaneously verify that detectors with internal dynamics can mitigate the impact.
The detectors we investigate are the $\chi^2$ detector and the multivariate exponentially weighted moving average (MEWMA) detector.
In addition to that, we investigate active noise injection into the controller dynamics to reveal otherwise stealthy attacks by preventing a successful completion of the first stage.

\citet{DetectorMetrics} did experiments with a water treatment system under both actuator and sensor attacks. Their results are related to the ones we obtain in this paper. 
However, \citet{DetectorMetrics} do not consider the first two stages of the attack, which are necessary for the stealthy execution of the attack. 
Although the controller with an integral action leads to a larger attack impact than a controller without an integral action both in our work as well as in \citep{DetectorMetrics}, we show that the integral action prevents the attack from being stealthy.
\citet{RealWorldDetectorComparisonACC19} compare four different detectors with respect to their attack detection capability in simulation and real-world experiments. 
In the present work, we will only compare two of the four detectors investigated in \citep{RealWorldDetectorComparisonACC19}, but look at a more sophisticated attack strategy. 
We also use a technique similar to the watermarking in \citep{RealWorldDetectorComparisonACC19} to achieve a detection of stealthy attacks. 
However, instead of injecting an additive noise signal to the controller output, $u(k)$, we inject the noise directly into the controller dynamics in this work.

The remainder of the paper is structured as follows. 
In Section~\ref{sec:ClosedLoopSystem}, we introduce the components of the experimental setup, such as the hardware, and the controllers and the anomaly detectors investigated. 
The sensor attack strategy with its three stages is presented in Section~\ref{sec:SensorAttack}. 
Furthermore, Section~\ref{sec:SensorAttack} shows the experimental results for each attack stages and compares them with the theoretical results.
Defense mechanisms such as the choice of the controller, the choice of the detector, and the active noise injection are discussed and investigated in Section~\ref{sec:DefenseMechanisms}. 
Section~\ref{sec:Conclusion} concludes the paper.

\emph{Notation:} Let $x\in\mathbb{R}^{n}$ be an $n$-dimensional real-valued vector and $A\in\mathbb{R}^{m \times n}$ a real-valued matrix with $m$ rows and $n$ columns. 
The $p$-norm of a vector $x$ is denoted by $\|x\|_p=\sqrt[p]{\sum_{i=1}^{n}|x_i|^p}$, $x_i$ is the $i$th element of $x$ and $\|x\|_{\infty}=\max_i(|x_i|)$. 
Let $A\in\mathbb{R}^{m_1 \times n_1}$ and $B\in\mathbb{R}^{m_2 \times n_2}$ then $\mathrm{diag}(A,B)\in\mathbb{R}^{(m_1+m_2) \times (n_1+n_2)}$ is a block diagonal matrix with $A$ and $B$ as its block diagonal entries. Further, $I_n$ represents the $n$-dimensional identity matrix.
A Gaussian random variable $x$ with mean $\mu$ and covariance matrix $\Sigma$ is denoted as $x\sim\mathcal{N}(\mu,\Sigma)$.
\section{The closed-loop system}
\label{sec:ClosedLoopSystem}
\begin{figure}
\centering
\includegraphics[scale=0.35]{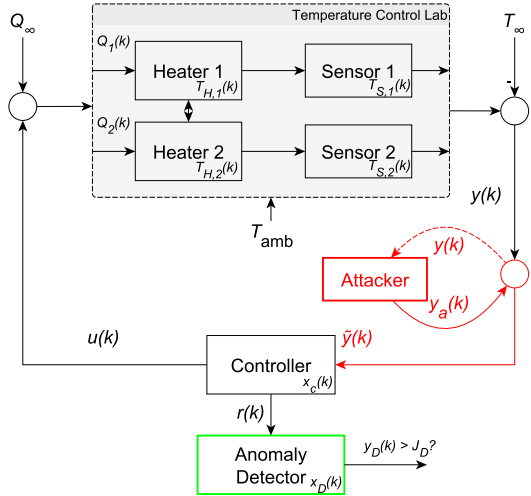}
\caption{Block diagram of the experimental setup depicting the TCLab, the attack on the measurements, the controller, and the anomaly detector, where the TCLab is controlled around the steady-state input~$Q_\infty$ and steady-state output~$T_{\infty}$. The attacker is able to eavesdrop on the measurements (dashed line) and inject a malicious additive signal $y_a(k)$ to the measurements.}
\label{fig:BlockDiagTCLab}
\end{figure}

In this section, we present  the closed-loop system composed of the temperature control lab, which is the hardware used in the experiment, the controller, and the anomaly detector.
Figure~\ref{fig:BlockDiagTCLab} shows a block diagram of the closed-loop system under a sensor attack.

\subsection{The Temperature Control Lab}
As the hardware in our experiment we use the Temperature Control Lab (TCLab) \citep{TCLabModelBenchmark}.
The TCLab is an Arduino-based process, which consists of two heaters with radiators and two temperature sensors, one for each heater (see Figure~\ref{fig:BlockDiagTCLab}).
The heaters are close to each other such that their temperatures are coupled. 
In this process, we can set the output power of each heater, $Q_1$ and $Q_2$, respectively, to a value between $0\,\%$ and $100\,\%$.
Different modelling strategies for the TCLab are evaluated by \citet{TCLabModelBenchmark} and the conclusion is that a four-dimensional physics-based model describes the process best, where the temperature unit for the dynamics is Kelvin while we use degree Celsius in the figures.
The first two states of the model represent the temperatures of the heater ($T_{H,1}$ and $T_{H,2}$) and their dynamics are based on convective and radiative heat transfers between the heaters and the ambient temperature $T_{\mathrm{amb}}$.
The last two states represent the state of each sensor ($T_{S,1}$ and $T_{S,2}$) and are represented by a linear low-pass filter.
A more detailed description of the TCLab and its dynamics can be found in Appendix~\ref{appendix:TCLab}.

\subsection{Controller design}
In this section, we design two controllers that control the TCLab around a certain steady state.
Here, we only provide a coarse description of the controller design, while more information on the controller design can be found in Appendix~\ref{appendix:TCLab}.
Remember that the main focus of this work lies on evaluating theoretical results on sensor attacks in an experiment with a real process. Therefore, we design the controllers such that they perform satisfactorily with respect to the steady state, but do not take any more constraints, such as rise time constraints, into account.

We design two linear controllers that control the TCLab at a steady-state temperature.
To find a linear model, we linearize the non-linear physics-based dynamics  around a steady-state value.
One can determine that in steady state, we have $T_{S,1}=T_{H,1}$ and $T_{S,2}=T_{H,2}$, so that we only need to find the steady-state values of the heater temperatures and the heater power. These are denoted as $T_{\infty}=[T_{H,1\infty},\, T_{H,2\infty}]^\top$ and $Q_{\infty}=[Q_{1\infty},\, Q_{2\infty}]^\top$ in Figure~\ref{fig:BlockDiagTCLab}.
The linearized dynamics of the TCLab and the equations to determine the steady-state values are provided in Appendix~\ref{appendix:TCLab}.

Since the experiments with the TCLab are conducted at room temperature, we set $T_{\text{amb}}=294.15\,\mathrm{K}=21\,{}^{\circ}\mathrm{C}$ and we choose $T_{H,\infty}=313.15\,\mathrm{K}=40\,{}^{\circ}\mathrm{C}$.
Linearizing the dynamics around this steady state and discretizing the linearized equations with a sampling time of $T_s=1\,\mathrm{s}$ leads to our linearized discrete-time model,
\begin{align*}
x(k+1)&=A x(k)+Bu(k),\\
y(k)&=Cx(k),
\end{align*}
where
\begin{align*}
A&=\begin{bmatrix} 
	0.9784  &  0.0113  &       0  &       0\\
    0.0113  &  0.9784  &       0  &       0\\
    0.0385  &  0.0002  &  0.9610  &       0\\
    0.0002  &  0.0430  &       0  &  0.9565\end{bmatrix},\\
B&=\begin{bmatrix} 
    0.0085  &  0\\
    0		&  0.0047\\
    0.0002  &  0\\
    0	  	&  0.0001\end{bmatrix},\ \mathrm{and}\ C=\begin{bmatrix}
0 & 0 & 1 & 0\\
0 & 0 & 0 & 1
\end{bmatrix}.
\end{align*}

Based on this linearized model, we want to design two different linear time-invariant output feedback controllers for the TCLab of the form
\begin{align*}
x_c(k+1)&=A_cx_c(k)+B_c\tilde{y}(k),\\
u(k)&=C_cx_c(k),
\end{align*}
where $x_c(k)\in\mathbb{R}^{n_c}$ is the internal controller state and $\tilde{y}(k)$ is the measurement received by the controller, where $\tilde{y}(k)=y(k)$ cannot be guaranteed due to the attacker (see Figure~\ref{fig:BlockDiagTCLab}).
The matrices $A_c$, $B_c$, and $C_c$ are matrices of appropriate dimension that we determine through the control design.
We design two different controllers because we want to investigate the influence of different controllers on the attack.
The controllers we design are an LQG controller and an LQG controller with integral action, subsequently called LQI controller.
For the control design, we need to choose cost matrices for both the state and the input, as well as covariance matrices for the process and measurement noise. 
We use the guidelines provided by \citet{LQGDesignPhilosophy} to determine the cost matrices for the LQR problem and also the covariance matrices for the Kalman filter. 

For the LQG controller, we choose state and control input cost matrices $Q_x=10I_{4}$ and $R_u=2I_2$, respectively, such that the controller minimizes $\sum_{k=0}^\infty 10x(k)^\top x(k)+2u(k)^\top u(k)$.
For the LQI controller, the system is extended by two integrator states, which ensure that $y(k)$ converges to the reference value of zero, since $y(k)$ represents the deviation of the output from the desired steady state.
Therefore, we choose state and control input cost matrices as $Q_{x,i}=\mathrm{diag}(10I_4,2I_{2})$ and $R_u=2I_2$, respectively, for the LQI controller such that the controller minimizes $\sum_{k=0}^\infty 10x(k)^\top x(k)+2x_{\mathrm{int}}(k)^\top x_{\mathrm{int}}(k)+2u(k)^\top u(k)$.
For the Kalman filter used in the LQG and LQI controllers, we set the process noise matrix as $\Sigma_w=5I_4$. 
Further, we choose the measurement noise matrix as $\Sigma_v=I_{2}$.
The matrices $A_c$, $B_c$, and $C_c$ are $A_c=A-BK-LC$, $B_c=L$, and $C_c=-K$, for the LQG controller, where $x_c(k)$ is an estimate $\hat{x}(k)$ of the state $x(k)$, $K$ is the controller gain, and $L$ is the steady-state Kalman filter gain.
Further, for the LQI controller, we have
\begin{align*}
A_c&=\begin{bmatrix}
A-BK_{\hat{x}}-LC & -BK_{\mathrm{int}}\\
0 & I_2
\end{bmatrix}, \\
B_c&=\begin{bmatrix}
L \\
-I_2
\end{bmatrix},\ 
C_c=\begin{bmatrix}
-K_{\hat{x}} & -K_{\mathrm{int}}
\end{bmatrix},
\end{align*}
and $x_c(k)=[\hat{x}(k)^\top\ x_{\mathrm{int}}(k)^\top]^\top$, where $\hat{x}(k)$ is again an estimate of the state $x(k)$ and $x_{\mathrm{int}}(k)$ is the integrator state. 
Here, $K_{\hat{x}}$ and $K_{\mathrm{int}}$ are the respective controller gains for the state estimate and the integrator to determine $u(k)$.
\begin{figure}
\centering
\includegraphics[width=0.5\textwidth]{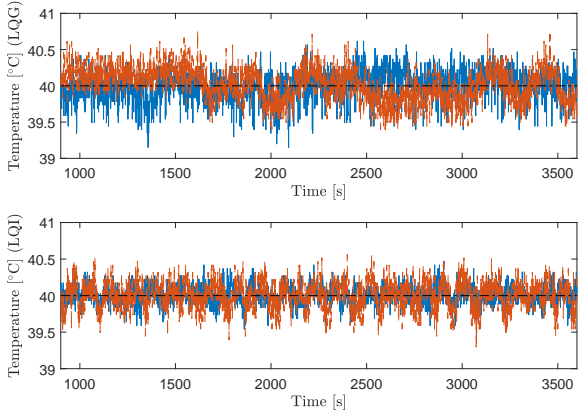}
\caption{The upper plot shows the steady-state behavior of the TCLab temperature measurements when the LQG controller is used, while the lower plot shows the steady-state behavior of the TCLab temperature measurements when the LQI controller is used. Both controllers manage to reach a steady-state close to $40\,{}^\circ\mathrm{C}$ and hold the temperatures there.}
\label{fig:NominalBehaviorLQGandLQI}
\end{figure}
In Figure~\ref{fig:NominalBehaviorLQGandLQI}, the measurement trajectories of the TCLab over a period of $2700\,\mathrm{s}$ are shown when the LQG controller is used (upper plot) and when the LQI controller is used (lower plot).
We do not show the first $900\,\mathrm{s}$ of the experiment, because they include the transient from room temperature to the desired temperature of $40\,{}^\circ\mathrm{C}$, which is indicated by a dash-dotted line in Figure~\ref{fig:NominalBehaviorLQGandLQI}.
We observe that both controllers are able to maintain the temperatures around the desired steady-state temperature, but the LQI controller is closer to the desired steady state due to the integral action.

\subsection{Anomaly detectors}
Both the LQG and the LQI controllers produce a residual signal, $\bar{r}(k)=y(k)-CT_c x_c(k)=y(k)-C\hat{x}(k)$, where $T_c\in\mathbb{R}^{n_x \times n_c}$ extracts the state estimate $\hat{x}(k)$ from the controller state.
The residual is the difference between the measured and the predicted output at time step $k$.
If our model for the TCLab is accurate around the steady state, the residual should be close to zero.
The sample mean $\mu_r$ and sample covariance $\Sigma_r$ of the residual in the nominal case shown in Figure~\ref{fig:NominalBehaviorLQGandLQI} are
\begin{align*}
\mu_r=\begin{bmatrix}
-0.1605\\
0.2905
\end{bmatrix}10^{-3}\ \mathrm{and}\ \Sigma_r=\begin{bmatrix}
0.0555  &  0.0013\\
0.0013  &  0.0482
\end{bmatrix}
\end{align*}
for the LQG controller and 
\begin{align*}
\mu_r=\begin{bmatrix}
-0.0005\\
-0.0222
\end{bmatrix}\ \mathrm{and}\ \Sigma_r=\begin{bmatrix}
0.0208  &  0.0015\\
0.0015  &  0.0426
\end{bmatrix}
\end{align*}
for the LQI controller.
This shows us that the controllers are predicting $y(k)$ with a high accuracy in steady state.
Hence, the residual signal can be used as a way to determine if the plant is behaving nominally or anomalously.
Therefore, we use the normalized residual $r(k)=\Sigma_r^{-\frac{1}{2}}\bar{r}(k)$ as the input to an anomaly detector\footnote{Typically, one would also incorporate the mean in the normalization, but since it is very small, we neglect the mean here.}, which produces an output $y_D(k)\in\mathbb{R}_{\geq 0}$.

The detector output is compared to a threshold $J_D>0$ and if the output exceeds the threshold, i.e., $y_D(k)>J_D$, an alarm is triggered and otherwise no alarm is triggered.
When choosing the threshold $J_D$ there is a trade-off between the false alarm rate and the ability to detect anomalies. 
Furthermore, \citet{DetectorMetrics} point out that there is also a trade-off between the number of false alarms and the impact of a stealthy attacker.
However, since neither the anomaly, nor the attacker, are here known to the operator, the threshold is often chosen as the smallest value that achieves an acceptable false alarm rate.

In this work, we want to investigate both a \emph{stateless} detector, which does not have internal dynamics, and a \emph{stateful} detector, which has internal dynamics, because the detector type will also influence the attacker's impact on the closed-loop system.
The stateless detector is a $\chi^2$ detector defined as
\begin{align*}
	y_D(k+1)=r(k)^\top r(k).
\end{align*}
Intuitively, this detector determines the size of the residual at the current time step and then compares it to a threshold.
\citet{RuthsMultivariate} give a closed-form solution for the $\chi^2$ detector threshold $J_D^{\chi^2}$ that achieves a desired average time between false alarms under the assumption that $r(k)\sim\mathcal{N}(0,I_2)$ and independent and identically distributed (i.i.d.).

The stateful detector is a MEWMA detector, defined as
\begin{equation}
\label{eq:MEWMADetector}
	\begin{aligned}
		x_D(k+1)&=\beta r(k)+(1-\beta)x_D(k)\\
		y_D(k+1)&=\frac{2-\beta}{\beta}\|x_D(k+1)\|_2^2,
	\end{aligned}
\end{equation}
where $\beta\in(0,1]$ and $x_D(0)=0$.
Further, if $y_D(k+1)>J_D$ then the detector state is reset to zero, i.e., $x_D(k+1)=0$.
The MEWMA detectors filters the residual signal through a low-pass filter, before it takes the squared Euclidean norm to determine the detector output.
For tuning the threshold of the MEWMA detector, $J_D^M$, no closed-form solution exists, but \citet{MEWMA_Markov} propose a way to approximate the threshold that achieves a certain average time between false alarms under the assumption that $r(k)\sim\mathcal{N}(0,I_2)$ and i.i.d.

For the thresholds, we choose $J_D=J_D^{\chi^2}=5.9915$ and $J_D=J_D^M=4.3918$ for the $\chi^2$ detector and the MEWMA detector with $\beta=0.2$, respectively. These thresholds result in an average time between false alarms of $20$ time steps based on the assumption that $r(k)\sim\mathcal{N}(0,I_2)$.
This assumption does typically not hold in reality.
Since tuning the threshold for general noise distributions is a non-trivial task and not the goal of this paper, we will use these thresholds from now on.
\section{The sensor attack model}
\label{sec:SensorAttack}
In this section, we introduce the sensor attack model. 
We begin by giving a short overview over the investigated sensor attack and its three stages. We also state the assumptions made on the attacker. 
Then we look into each of the stages of the attack in more detail and investigate how the theoretical results for each attack stage hold up in the experiment.
Before that we make the following assumption.
\begin{assumption}
\label{assum:SteadyStateBehavior}
The system has reached its steady state before the attack begins.
\end{assumption}
In control system, we often want to achieve a certain steady state. For example, to achieve an optimal temperature for a chemical reaction or operate a centrifuge at a desired speed. 
Therefore, we assume that the system is in its steady state, when the attacker attacks.
Furthermore, steady-state behavior is a common assumption in the literature on security of control systems.
\subsection{Attacker model}
Let us begin with a short overview on the attack procedure and introduce the attacker model.
\begin{assumption}
\label{assum:AttackerKnowledge}
The attacker has full model knowledge about the linearized plant and controller dynamics and, therefore, knows the matrices $A$, $B$, $C$, $A_c$, $B_c$, $C_c$, $T_c$, and the covariance matrices, $\Sigma_w$ and $\Sigma_v$, assumed by the operator. 
Furthermore, the attacker knows the detector used, its threshold $J_D$, and the covariance matrix of the residual, $\Sigma_r$.
Moreover, the attacker is able to read and manipulate the measurements $y(k)$ from $k\geq k_{\mathrm{I}}\geq 0$ on.
\end{assumption}

Assumption~\ref{assum:AttackerKnowledge} shows that we consider a powerful attack that does not only have perfect knowledge about the plant, controller, and detector model used but is also able to manipulate and eavesdrop on the sensor measurements.
This is in line with \citet{SecrecySystemsShannon} who argues that a system should be designed for the worst case and given enough time the attacker is able to obtain a perfect model of the plant and controller.
Further, note that without loss of generality we set the time when the attacker enters the closed-loop system to $k=k_{\mathrm{I}}$.
Furthermore, due to the ability to manipulate the sensor measurements, the controller receives $\tilde{y}(k)=y(k)+y_a(k)$ instead of $y(k)$ as shown in Figure~\ref{fig:BlockDiagTCLab}, where $y_a(k)$ is a signal constructed by the attacker.
Finally, observe that the attacker does not have access to real-time data from inside the controller and the detector, $x_c(k)$ and $x_D(k)$, respectively

\begin{assumption}
\label{assum:AttackerStealthiness}
The attacker wants to remain stealthy to the detector, i.e., $y_D(k+1)\leq J_D$ once the attacker has started to inject a non-zero attack signal $y_a(k)$.
\end{assumption}
A detection of the attack by the system operator leads to counter measures against the attack, which is why the attacker wants to remain stealthy. Furthermore, if the attacker is only eavesdropping and has not yet injected a non-zero $y_a(k)$ the attack is impossible to detect by using the proposed anomaly detector.

Due to Assumption~\ref{assum:AttackerStealthiness}, the attacker uses the following attack strategy \citep{RuthsMultivariate},
\begin{align}
\label{eq:AttackStrategy}
y_a(k)=-y(k)+CT_cx_c(k)+\Sigma^{-\frac{1}{2}}_r a(k),
\end{align}
where $T_cx_c(k)=\hat{x}(k)$ is the estimate of the plant's state made by the operator.
This attack strategy, if successfully executed, gives the attacker full control over the residual signal, i.e., $r(k)=a(k)$, and, therefore, makes it possible for the attacker to remain stealthy.
The attack \eqref{eq:AttackStrategy} is a closed-loop attack, since it uses the measurements directly for the attack signal and $a(k)$ can be seen as a reference signal that the attacker injects into the system. 
However, for a successful execution of \eqref{eq:AttackStrategy}, the attacker,  in addition to Assumption~\ref{assum:AttackerKnowledge},  needs to know $CT_cx_c(k)$, which cannot be known immediately when the attacker enters the system.
Therefore, the first stage, \textbf{Stage~I}, of the attack strategy is to estimate the quantity $CT_cx_c(k)$ perfectly.
This already shows us that although the attacker is powerful according to Assumption~\ref{assum:AttackerKnowledge}, it cannot immediately launch \eqref{eq:AttackStrategy} when entering the system.
Note that $y_a(k)=0$ in Stage~I such that the attacker is only eavesdropping and does not need to be concerned with its stealthiness.

Once the attacker has managed to estimate $CT_cx_c(k)$ it can launch the attack in \eqref{eq:AttackStrategy} and take over full control of the detector input.
Due to Assumption~\ref{assum:AttackerStealthiness} the attacker needs to design $a(k)$ in such a way that no alarm is triggered. 
In case the detector has an internal state $x_D(k)$, the attacker needs to estimate $x_D(k)$ as well. 
Otherwise, a certain detector state in combination with the input $a(k)$ can trigger an alarm and the attacker will also be able to have a larger stealthy impact if it has knowledge of the detector state.
Therefore, \textbf{Stage~II} of the attack is to estimate the detector state.
Finally, when both $CT_cx_c(k)$ and the detector state are known to the attacker, it can design a trajectory for $a(k)$ that maximizes the attack impact, which is defined in Section~\ref{sec:StageIII}, while remaining stealthy. Injecting this attack trajectory is \textbf{Stage~III} of the attack.

Next, we describe each of the three stages of the attack in more detail by presenting theoretical results for each stage and evaluate the results with our experimental setup.
For the experiment, we assume that a MEWMA detector \eqref{eq:MEWMADetector} is used to be able to illustrate all three stages of the attack.
As in Figure~\ref{fig:NominalBehaviorLQGandLQI}, we do not show the first $900\,\mathrm{s}$, because after that time the system has reached its steady state.
In addition to that, the results we show in each of the stages for the LQG or LQI controller are from the same experiment for the respective controller.
This, for example, means that the controller state estimate of Stage~I and the detector state estimate of Stage~II will be used in Stage~III.
Furthermore, we will denote the start time and length of each stage by $k_i$ and $N_i$, respectively, where $i\in\lbrace I,II,III\rbrace$.
Since the stages follow up on each other we have, $k_{\mathrm{I}}=900$, $k_{\mathrm{II}}=k_{\mathrm{I}}+N_{\mathrm{I}}-1$, and $k_{\mathrm{III}}=k_{\mathrm{I}}+N_{\mathrm{I}}+N_{\mathrm{II}}-1$.

\subsection{Stage~I: Controller state estimation}
\label{sec:StageI}
\begin{figure}
\centering
\includegraphics[width=0.5\textwidth]{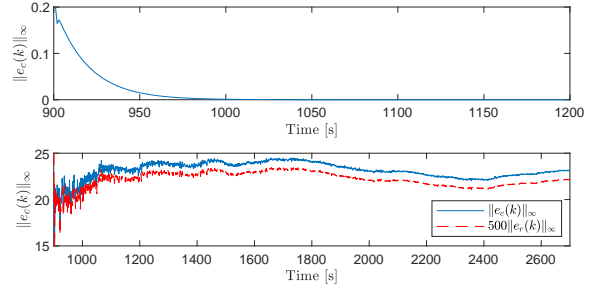}
\caption{The maximum controller state estimation error $e_c(k)$ of the attacker during Stage~I is shown in this figure. For the LQG controller case (upper plot) the estimation procedure is run for $300\,\mathrm{s}$, while for the LQI controller case (lower plot) the estimation procedure is run for $1800\,\mathrm{s}$.}
\label{fig:StageIforLQGandLQI}
\end{figure}

In the first stage of the attack, the attacker needs to obtain a perfect estimate of the operator's predicted measurement, since otherwise it cannot launch the attack \eqref{eq:AttackStrategy}.
Therefore, the attacker wants to determine an estimate, $\hat{x}_c(k)$, of the controller state, $x_c(k)$, such that $\lim_{k\rightarrow\infty}\|x_c(k)-\hat{x}_c(k)\|_2=0$. 
We provided a necessary and sufficient condition for when the perfect estimation is possible in \citep{UmsonstAutomatica21}.
\begin{thm}[\citet{UmsonstAutomatica21}]
	\label{thm:StageI}
		Let the plant and controller dynamics be linear and the plant be influenced by Gaussian process and measurement noise. Then under Assumption~\ref{assum:AttackerKnowledge} and with knowledge of $\Sigma_w$ and $\Sigma_v$, the attacker can perfectly estimate the controller state $x_c(k)$ if and only if $\rho(A_c)\leq 1$, where $A_c$ is the controller's system matrix.
\end{thm}
In addition to Theorem~\ref{thm:StageI}, we point out that if $\rho(A_c)<1$ holds then the estimation is exponentially fast, and the attacker can use a non-optimal time-invariant observer and does not need to know the noise properties.
As mentioned before, Theorem~\ref{thm:StageI} was derived under the assumption of a linear closed-loop system under the influence of Gaussian noise.
Therefore, we now want to determine if the result still holds in our experimental setup, where the plant dynamics are non-linear and the noise has an unknown distribution.
Let $e_c(k)=x_c(k)-\hat{x}_c(k)$ such that $\|e_c(k)\|_{\infty}$ denotes the maximum absolute estimation error of the controller state by the attacker.
If we use the LQG controller, we have $\rho(A_c)=0.9449<1$ such that the attacker is able to use a non-optimal observer to estimate the detector state exponentially fast.
For the LQG case, the attacker uses an open-loop estimator of the form
\begin{align}
\label{eq:OpenLoopStageI}
\hat{x}_c(k+1)=A_c\hat{x}_c(k)+B_cy(k),
\end{align}
which guarantees an exponentially fast convergence to the true detector state.
Therefore, Stage~I in the LQG case is executed for $300\,\mathrm{s}$ in the experiment and $\|e_c(k)\|_{\infty}$ is depicted in the upper plot of Figure~\ref{fig:StageIforLQGandLQI}.
We see that the convergence is indeed exponentially fast and after $300\,\mathrm{s}$ the estimation error is smaller than $2.3\cdot 10^{-8}$. 
Although the estimation error is not zero, we consider that the attacker has successfully estimated the controller state after Stage~I in the LQG controller case.
Next, we look at the LQI controller case, where we have $\rho(A_c)=1$ due to the integral action. 
Therefore, the attacker cannot use the open-loop estimator \eqref{eq:OpenLoopStageI} and needs to use a closed-loop estimate, which takes the plant dynamics and noise statistics into account as well.
With perfect system knowledge the attacker can design a time-varying Kalman filter to estimate the controller state perfectly. 
However, in our case, the attacker does not have perfect knowledge about the noise processes.
For example, the covariance matrix of the process noise, $\Sigma_w$, is chosen as $\Sigma_w=5I_4$ to obtain a certain control performance, but it is not guaranteed to be the correct covariance matrix for the process noise.
We already see that the assumptions necessary for Theorem~\ref{thm:StageI} do not hold in the experiment and the attacker is not able to design an optimal time-varying Kalman filter to estimate $x_c(k)$.
Nevertheless, we want to investigate how the time-varying Kalman filter performs in estimating $x_c(k)$ for the chosen noise covariance matrices.
Due to the integral action, exponential convergence is not guaranteed either and, therefore, Stage~I is executed for $N_{\mathrm{I}}=1800\,\mathrm{s}$ in the LQI controller case.
The maximum estimation error is shown as the solid line in the lower plot of Figure~\ref{fig:StageIforLQGandLQI}.
The trajectory of the maximum estimation error in the LQI controller case is very different from the trajectory in the LQG controller case.
For example, the maximum absolute error has not converged to zero after $1800\,\mathrm{s}$ and the error even increase again towards the end of Stage~I.
This shows us that the attacker is not able to obtain a perfect estimate of the controller state in our experiment.
However, the attack \eqref{eq:AttackStrategy} does not need a perfect estimate of the complete controller state, but only a perfect estimate of the residual.
Hence, we take a look at the estimation error of the residual signal, $e_r(k)=r(k)-\hat{r}(k)=\Sigma_r^{-\frac{1}{2}}(CT_c\hat{x}_c(k)-CT_cx_c(k))$, as well.
In the lower plot of Figure~\ref{fig:StageIforLQGandLQI} the maximum residual estimation error $\|e_r(k)\|_{\infty}$ scaled with a factor of $500$ is shown by the dashed line.
Here, we observe that $\|e_r(k)\|_{\infty}\in[0.03,0.05]$ for all of Stage~I.
Because of this observation, we want to investigate if the stealthy attack is still possible and also consider the LQI controller case in Stage~II and Stage~III in the following.

We want to point out that the controller state estimation continues throughout Stage~II and Stage~III, because the attacker needs the true controller state to launch \eqref{eq:AttackStrategy}, but it does not have access to the controller state itself. 
Therefore, the estimation needs to continue throughout the whole attack sequence.

\subsection{Stage~II: Detector state estimation}
\label{sec:StageII}
Assume now that the attacker successfully managed to estimate the controller state. Then it can launch the attack in \eqref{eq:AttackStrategy}, which gives the attacker full control over the detector input $r(k)$.
However, if the detector has an internal state, $x_D(k)$, which can, for example, be reset, an inappropriately chosen $a(k)$ can trigger an alarm and lead to the detection of the attacker.
Therefore, in Stage~II, the attacker tries to find an estimate, $\hat{x}_D(k)$, such that $\|x_D(k)-\hat{x}_D(k)\|_2\rightarrow 0$ as $k\rightarrow\infty$. 
Due to the reset in the MEWMA detector, the attacker injects a carefully designed signal $a(k)$ into the detector, which does not trigger an alarm and causes the estimate $\hat{x}_D(k)$ to converge to the true state.
We determine the following result on the length, $N_{\mathrm{II}}$, of Stage~II to determine a certain accuracy $\gamma$ for the detector state estimation error $e_D(k)=x_D(k)-\hat{x}_D(k)$.
\begin{proposition}[\citet{UmsonstACC19}]
\label{prop:StageII}
Assume that the attack executes its strategy \eqref{eq:AttackStrategy} such that $r(k)=a(k)$.
If the MEWMA detector is used, then ${\|e_D(k)\|_2\leq \gamma}$ after $N_{\mathrm{II}}=\frac{\ln\left(\frac{\sqrt{2-\beta}\gamma}{\sqrt{\beta J_D^M}}\right)}{\ln(1-\beta)}$ time steps independent of the value of $x_D(k_{\mathrm{II}})$.
\end{proposition}
In \citep{UmsonstACC19}, we show that during Stage~II the attacker is able to simultaneously estimate the detector state and inject a stealthy attack signal, which increases the expected value of the operator's estimation error, $\mathbb{E}\lbrace e(k)\rbrace=\mathbb{E}\lbrace x(k)-\hat{x}(k)\rbrace$ while mimicking the statistics of the detector output. 
The dynamics of $\mathbb{E}\lbrace e(k)\rbrace$ under the attack \eqref{eq:AttackStrategy} are
\begin{align*}
\mathbb{E}\lbrace e(k+1)\rbrace=A\mathbb{E}\lbrace e(k)\rbrace-L\Sigma_r^{\frac{1}{2}}a(k)
\end{align*}
and we assume $\mathbb{E}\lbrace e(k_{\mathrm{II}})\rbrace=0$ at the beginning of Stage~II.
The attacker can design a stealthy attack signal that mimics the detector output statistics by solving the following optimization problem each time step,
\begin{align*}
&\max_{a(k)} \|\mathbb{E}\lbrace e(k+1)\rbrace \|_{\infty}\\ 
&\mathrm{s.t.}\ \frac{2-\beta}{\beta}\|\beta a(k)+(1-\beta)\hat{x}_D(k)\|_2^2=y_{D,k+1},
\end{align*}
where $k\geq k_{\mathrm{II}}$, $y_{D,k+1}$ is a sample from a distribution with support $[0,J(k+1)]$ that mimics the statistics of the detector output without triggering an alarm, and $J(k)= (1-(1-\beta)^k)J_D^M\leq J_D^M$. 
Furthermore, it is assumed that $r(k)\sim\mathcal{N}(0,I_2)$, when drawing the sample $y_{D,k+1}$.
Note that since $\mathbb{E}\lbrace e(k_{\mathrm{II}})\rbrace=0$, the attacker knows all relevant signals for this optimization problem and can also solve it offline before the attack.
More details can be found in \citep{UmsonstACC19}.

\begin{figure}
\centering
\includegraphics[width=0.5\textwidth]{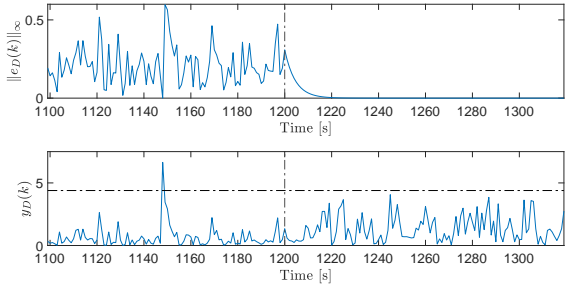}
\caption{For the LQG controller case, the maximum detector state estimation error is shown in the upper plot before and during Stage~II of the attack. We observe that the estimation error decreases exponentially fast during Stage~II. The lower plot shows the detector output before and during Stage~II. During Stage~II the output is still stochastic and never crosses the threshold, which comes from the attack design during Stage~II.}
\label{fig:StageIIforLQG}
\end{figure}

From Proposition~\ref{prop:StageII} we determine that $\|x_D(k)-\hat{x}_D(k)\|\leq 1.6406\cdot 10^{-12}$, when $N_{\mathrm{II}}=120$, $\beta=0.2$ and $J_D^{M}=4.3918$. Therefore, we let Stage~II run for $120\,\mathrm{s}$ for both the LQG and the LQI controller case.
The results of Stage~II for the LQG controller case are shown in Figure~\ref{fig:StageIIforLQG}, where we also included the last $100\,\mathrm{s}$ before the start of Stage~II and the vertical dash-dotted line marks the beginning of Stage~II.
The upper plot shows the maximum detector state estimation error, $\|e_D(k)\|_{\infty}$, where $e_D(k)=x_D(k)-\hat{x}_D(k)$.
We observe that before the start of Stage~II the maximum detector state estimation error fluctuates a lot, since in this time period $\hat{x}_D(k)=0$. 
As soon as the attacker starts Stage~II with its controller estimate from Stage~I, $\|e_D(k)\|_{\infty}$ decreases exponentially fast to zero and we have that $\|e_D(k)\|_{\infty}\leq 9.06\cdot 10^{-13}$ at the end of Stage~II.
The lower plot of Figure~\ref{fig:StageIIforLQG} shows the MEWMA detector output before and during Stage~II.
Recall that the attack signal is designed in such a way that it mimics the detector output statistics under the assumption that the residual has a standard Gaussian distribution while not crossing the threshold.
First, we note that the attacker successfully manages to remain stealthy and the detector output does not cross the threshold.
Second, the detector output is still stochastic, but from a visual inspection it seems noisier during Stage~II than before.
The reason for that is that when designing the attack the assumption that $r(k)\sim\mathcal{N}(0,I_2)$ does typically not hold in the experiment.
This noisier behavior could make an observant system operator suspicious.

\begin{figure}
\centering
\includegraphics[width=0.5\textwidth]{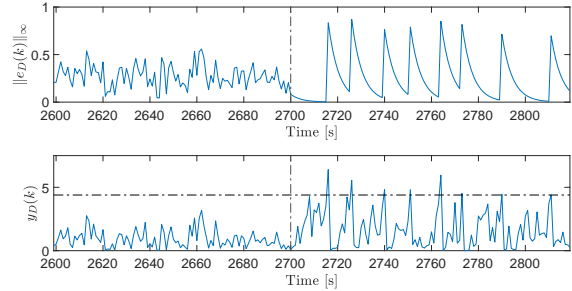}
\caption{For the LQI controller case, the maximum detector state estimation error is shown in the upper plot before and during Stage~II of the attack. We observe that the estimation error sharply increases at the time instances where the detector output crosses the threshold (lower plot). The sharp increase in the estimation error is caused by the reset of the detector state when the threshold is exceeded.}
\label{fig:StageIIforLQI}
\end{figure}

Next, we consider Stage~II of the LQI controller case. Recall that the attacker did not manage to successfully complete Stage~I, but has a small estimation error for the part of the controller state that is needed to launch \eqref{eq:AttackStrategy}.
Therefore, it will not have complete control over the input to the detector as in the case of the LQG controller.
This uncertainty leads to very different outcomes than in the LQG controller case as seen in Figure~\ref{fig:StageIIforLQI}.
The upper plot in Figure~\ref{fig:StageIIforLQI} shows again the maximum estimation error before and during Stage~II, where the beginning of Stage~II is marked by the vertical dash-dotted line.
During Stage~II, we observe that the estimation error initially decreases, but then increases in a spike just to decrease again. 
Further, the detector state estimation error at the end of Stage~II did not decrease to zero and the maximum estimation error equals $0.1171$.
The spikes in the detector state estimation error can be explained by looking at the detector output during Stage~II (lower plot in Figure~\ref{fig:StageIIforLQI}). 
We observe that the detector output in Stage~II crosses the threshold, which leads to a reset of the detector state. 
This reset explains the sharp increase of the maximum detector state estimation error.
Recall that the attacker does neither have access to $x_D(k)$ nor to $y_D(k)$, so it does not know when the reset is actually happening.
So we see that in the LQI controller case the uncertainty in the attacker's controller state estimate leads to a detection of the attacker during the second stage of the attack.

\subsection{Stage~III: Targeting the plant}
\label{sec:StageIII}
In the third and final stage of the attack, the attacker launches its stealthy attack that maximizes the impact on the plant while remaining stealthy.
Here, the attacker wants to maximize the expected value of the plant's state $\mathbb{E}\lbrace x(N_a)\rbrace$ at the end of attack Stage~III, where $N_a=k_{I}+N_{\mathrm{I}}+N_{\mathrm{II}}+N_{\mathrm{III}}$ marks the end of the third stage of the attack.
Let $a\in\mathbb{R}^{N_{\mathrm{III}}n_y}$ be the complete attack trajectory during Stage~III, then $\mathbb{E}\lbrace x(N_a)\rbrace=T_{xa}a$ under the assumption of linear dynamics for both the plant and the controller, where $T_{xa}\in\mathbb{R}^{n_x\times N_{\mathrm{III}}n_y}$ describes the influence of the attack trajectory on the expected plant's state at the end of the attack.
In \citep{UmsonstACC17}, we pose a convex optimization problem that lets us estimate the worst-case attack impact of a stealthy attack.
Under the assumption of linear plant and controller dynamics the problem of worst-case impact estimation is formulated as
\begin{align}
	\label{eq:StageIIIWorstCaseImpact}
	\max \|T_{xa}a\|_{\infty}\quad \mathrm{s.t.}\quad y_D(k+1)\leq J_D,
\end{align} 
where $k\in\lbrace k_{\mathrm{III}},\ldots, k_{\mathrm{III}}+N_{\mathrm{III}}-1\rbrace$.
Note that this optimization problem assumes that at the beginning of the third stage of the attack $x(k)=0$, $x_c(k)=0$, and $x_D(0)=0$, if the detector has an internal state. 
Due to Assumption~\ref{assum:SteadyStateBehavior}, assuming $x(k)=0$ and $x_c(k)=0$ is not a strict assumption, because in steady state these values should be close to zero.
Furthermore, the change in the states due to Stage~II is typically small, because of the short time during which Stage~II is executed.
Moreover, since at the beginning of Stage~III we assumed $x_D(k)=0$ when determining the worst-case impact, we set the first attack signal in Stage~III as
\begin{align*}
y_a(k)=-y(k)+CT_cx_c(k)+\Sigma^{-\frac{1}{2}}_r a(k)-\frac{1-\beta}{\beta}\hat{x}_D(k),
\end{align*}
which removes the influence of the detector state at the beginning of the third stage of the attack, so that this assumption is fulfilled.
Finally, we set the attack length for Stage~III to $N_{\mathrm{III}}=1800\,\mathrm{s}$, which is equivalent to a worst-case attack that lasts for half an hour.
With that, we want to evaluate if the impact on the real plant in the experiment, the TCLab, is the same as the theoretical estimation of the worst-case attack impact provided by \eqref{eq:StageIIIWorstCaseImpact}.
\begin{figure}
\centering
\includegraphics[width=0.5\textwidth]{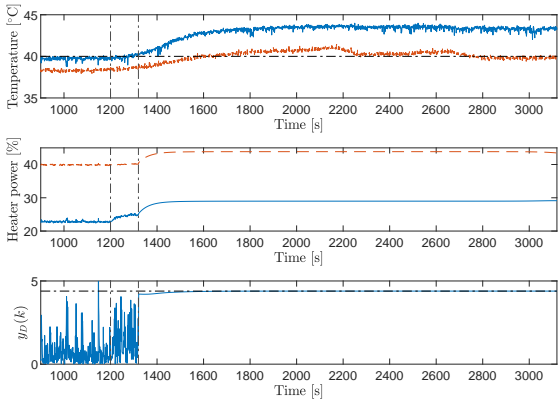}
\caption{This figure shows several closed-loop system trajectories before Stage~II, during Stage~II and during Stage~III of the attack for the LQG controller case. The beginning of Stage~II is marked by the first vertical dash-dotted line from the left and the beginning of Stage~III is marked by the second vertical dash-dotted line. The upper plot shows the two measurements taken from the TCLab with the desired steady-state temperature as a horizontal dash-dotted line, the center plot shows the power applied to the heaters, and the lower plot shows the detector output with the threshold as a horizontal dash-dotted line. These trajectories show that the attacker is able to launch a stealthy attack.}
\label{fig:StageIIIforLQG}
\end{figure}

We begin with the LQG controller case, for which the attacker could successfully complete Stage~I and Stage~II. Therefore, the attacker has now all the information needed to launch a worst-case attack.
When solving the worst-case impact estimation problem \eqref{eq:StageIIIWorstCaseImpact} under the assumption of linearity, we obtain that the theoretical maximum increase in temperature due to the attack is $4.79\,{}^\circ\mathrm{C}$ and the attack will focus on the temperature corresponding to the first measurement.
The measurements of the TCLab, the input to the TCLab, and the detector output $300\,\mathrm{s}$ before Stage~II, during Stage~II, and during Stage~III are shown in Figure~\ref{fig:StageIIIforLQG}.
The lower plot of Figure~\ref{fig:StageIIIforLQG} shows the detector output during the attack and we observe that the attacker is able to remain stealthy during the attack, since no alarm is triggered. 
However, we also observe that the detector output approaches a constant value equal to the threshold in Stage~III.
Therefore, the attack itself is easily detected by a visual inspection of the detector output. 
The reason for this constant detector output is that the constraints in the worst-case impact estimation problem \eqref{eq:StageIIIWorstCaseImpact} only enforce stealthiness, and not, for example, mimicry of the detector output statistics.
Furthermore, by looking at the input to the plant (center plot in Figure~\ref{fig:StageIIIforLQG}), we note that the attack in Stage~III will lead to a constant input to the plant, such that it reaches a new steady state.
We also observe that in Stage~II the input also increased slightly, due to the attack design, which maximizes the operator's estimation error. 
Stage~II is, however, too short to have a significant impact in the LQG controller case.
Finally, the outputs of the TCLab are shown in the upper plot of Figure~\ref{fig:StageIIIforLQG}. 
As expected the attack targets the temperature corresponding to the first measurement (solid line).
The temperature in that heater increases from an average of $39.73\,{}^\circ\mathrm{C}$ before the attack to an average of $43.37\,{}^\circ\mathrm{C}$. 
This is an increase of $3.64\,{}^\circ\mathrm{C}$, which is approximately $1.15\,{}^\circ\mathrm{C}$ smaller than the estimated worst-case attack impact.

Next, we present the attack for Stage~III when an LQI controller is used. Since the attack in Stage~II is based on samples of the truncated detector output distribution and the attacker could theoretically remain undetected in Stage~II.
When solving \eqref{eq:StageIIIWorstCaseImpact} in the LQI controller case, we obtain that the theoretical worst-case impact is a change of $9.13\,{}^\circ\mathrm{C}$ from the steady state before the attack and the attacker targets the temperature of the second heater in this case.
This demonstrates already two things before looking at the results from the experiment. 
First, the attacker will target a different state depending on the controller used.
Second, the use of an LQI controller increases the theoretically possible worst-case impact by almost a factor of two\footnote{Note that the optimization problem \eqref{eq:StageIIIWorstCaseImpact} uses the estimated residual covariance matrix from the nominal LQI controller case (lower plot in Figure~\ref{fig:NominalBehaviorLQGandLQI}), so it is different from the covariance matrix used in the LQG case. 
However, if we would use the covariance matrix estimated from the LQG case, the estimated impact would be even larger.}. 
Intuitively, if the attacker feeds a constant signal into the LQG controller, the controller output converges to a constant signal as well, while the LQI conroller will integrate the constant signal up, such that the impact increases compared to the LQG controller.

\begin{figure}
\centering
\includegraphics[width=0.5\textwidth]{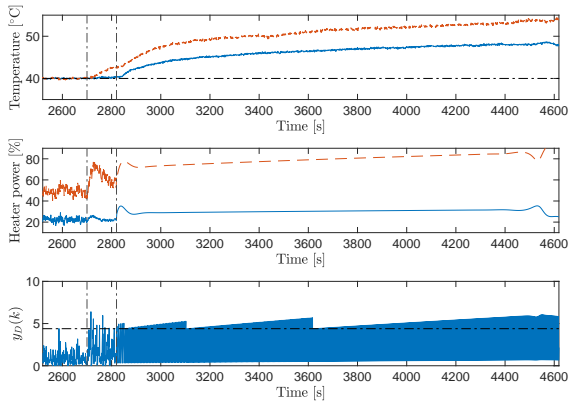}
\caption{This figure shows several closed-loop system trajectories before Stage~II, during Stage~II and during Stage~III of the attack for the LQI controller case. The beginning of Stage~II is marked by the first vertical dash-dotted line from the left and the beginning of Stage~III is marked by the second vertical dash-dotted line. The upper plot shows the two measurements taken from the TCLab with the desired steady-state temperature as a dash-dotted horizontal line, the center plot shows the power applied to the heaters, and the lower plot shows the detector output with the threshold as a horizontal dash-dotted line. These trajectories show that the attacker is not able to remain stealthy, but has a larger impact on the system than in the LQG case.}	
\label{fig:StageIIIforLQI}
\end{figure}

The trajectory of the true TCLab measurements, the control input, and the detector output $300\,\mathrm{s}$ before Stage~II, during Stage~II, and during Stage~III are shown in Figure~\ref{fig:StageIIIforLQI}.
We start by investigating the temperature increase during Stage~III shown in the upper plot of Figure~\ref{fig:StageIIIforLQI}. 
We observe that the temperature of the second heater increases constantly from an average of $40\,{}^\circ\mathrm{C}$ before the attack to $54.06\,{}^\circ\mathrm{C}$ at the end of the attack.
This is a temperature increase of $14.06\,{}^\circ\mathrm{C}$, which is approximately $5\,{}^\circ\mathrm{C}$ larger then the theoretical worst-case impact.
Part of this increase compared to the theoretical value is also due to the attack during Stage~II. 
When Stage~III starts the temperature has already risen from $40\,{}^\circ\mathrm{C}$ to around $42.5\,{}^\circ\mathrm{C}$. Discounting that increase the experimental impact is only approximately $2.5\,{}^\circ\mathrm{C}$ larger then the theoretical worst-case impact.
Other than in the LQG controller case, the actual attack impact is larger than but still close to the estimated one.
The center plot of Figure~\ref{fig:StageIIIforLQI} shows the trajectory of the input to the TCLab.
We note that the input during Stage~III seems to be linearly increasing, which is a combination of an almost constant attack signal at the beginning of Stage~III and the integral action in the controller.
We also observe the large increase in the control input during Stage~II, which could explain the crossings of the detector threshold during Stage~II.
While the impact of the attack in the LQI controller case is more severe compared to the LQG case, we would like to remind the reader that this attack can already be detected in Stage~II.
Furthermore, the detector output has an oscillating behavior in Stage~III, where it repeatedly crosses the threshold and is reset to zero (see lower plot in Figure~\ref{fig:StageIIIforLQI}).
Therefore, we think that even if the attacker managed to stealthily execute Stage~II, its attack strategy in Stage~III would lead to a quick detection.

\section{Discussion of defense mechanisms}
\label{sec:DefenseMechanisms}
In this section, we want to discuss the results we obtained from the experiments and how the choice of the controller and the detector can be used as a defense mechanism.
Furthermore, we investigate active noise injection into the controller dynamics to prevent controller state estimation.

\subsection{Choice of controller and detector}
We begin by discussing how the choice of controller affects the sensor attack.
From Theorem~\ref{thm:StageI}, we know that a controller with poles inside or on the unit circle enables the attacker to estimate the controller state perfectly. 
However, this result is obtained under the assumption of linear plant dynamics and Gaussian noises.
In Section~\ref{sec:StageI}, we showed that in the LQG controller case the attacker is able to estimate the controller state exponentially fast, but in the LQI controller state it is not clear if the attacker can perfectly estimate the controller state. Even after a six times longer time horizon for Stage I than in the LQG controller case, the estimation error is not close to zero.
The lack of exact knowledge of the controller state leads to a detection by the detector in later stages of the attack as shown in Figure~\ref{fig:StageIIforLQI} and Figure~\ref{fig:StageIIIforLQI}.
Therefore, we conclude that if a controller with stable dynamics is used the attacker is able to execute all three stages of the attack without being detected. If the controller has eigenvalues on the unit circle, this is not possible and will lead to a quick detection.
Luckily, controllers in practice often include integral action to take care of disturbances and reach a desired steady state.
Hence, the integral action of a controller can lead to a better protection of the system.
However, if the attacker manages via another way to obtain a perfect estimate of the controller state the worst-case stealthy attack impact in the LQI controller case can be much larger than in the LQG controller case.
In addition to that, the experimental results in \citet{DetectorMetrics} also show that having an integral part makes sensor attacks worse, but can help mitigate actuator attacks. 
A theoretical insight into why if an integrator is present in the controller, sensor attacks have a large impact, while actuator attacks are mitigated is provided by \citet{IntegratorAndSensorAttackHenrik2021}.
Therefore, it is always important to consider several attack strategies, when evaluating the resilience of the closed-loop system against attacks.

Next, we want to investigate how the choice of detector influences the attack impact.
A metric for detector comparison, which takes the attack impact and the time between false alarms into account, is proposed by \citet{DetectorMetrics}.
A detector is considered better than another detector if using one detector results in a lower stealthy attack impact than using another detector, while guaranteeing the same amount of false alarms.
Therefore, the detector choice can also be seen as a defense mechanism.
Here, we will compare the attack impact on the TCLab when an LQG controller and a $\chi^2$ detector are used, with the attack impact when a MEWMA detector is used. Recall that both detector thresholds are tuned to achieve the same average time between false alarms under the assumption of $r(k)\sim\mathcal{N}(0,I_2)$.
Since the LQI controller prevents the attacker from a stealthy attack, we will not consider the case of an LQI controller in this comparison.

\begin{figure}
\centering
\includegraphics[width=0.5\textwidth]{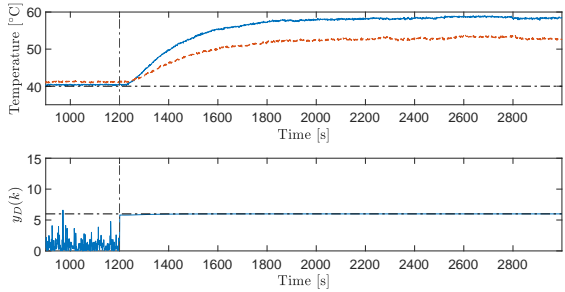}
\caption{This figure shows several closed-loop system trajectories during Stage I and during Stage III of the attack for the LQG controller case when the feedback loop is equipped with a $\chi^2$ detector. The beginning of Stage III is marked by the vertical dash-dotted line. The upper plot shows the two measurements taken from the TCLab with the desired steady-state temperature as a horizontal dash-dotted line, and the lower plot shows the detector output with the threshold as a horizontal dash-dotted line. These trajectories show that the attacker is able to launch a stealthy attack.}
\label{fig:StageIIIforLQGChi2}
\end{figure}

The $\chi^2$ detector does not have an internal state such that the attacker does not need to execute Stage II of the attack.
Therefore, one could argue that the MEWMA detector has already an advantage over the $\chi^2$ detector.
However, the execution of Stage II does not take a long time, $N_{\mathrm{II}}=120\,\mathrm{s}$ in Figure~\ref{fig:StageIIforLQG}. 
Hence, the extra stage does not seem to be a big inconvenience for the attacker.
Figure~\ref{fig:StageIIIforLQGChi2} shows trajectories of the true measurements of the TCLab, of the input to the TCLab, and of the detector output for Stage I and Stage III when an LQG controller and a $\chi^2$ detector are used.
As in the case with a MEWMA detector, the attacker will target the first heater in its attack, when solving \eqref{eq:StageIIIWorstCaseImpact}, and the theoretical impact is $16.7881\,{}^\circ\mathrm{C}$, which is around $3.5$ times larger than when a MEWMA detector is used.
This already shows us that under the $\chi^2$ detector the attacker is able to launch a stronger attack.
However, other than in the MEWMA case, the actual impact on the TCLab is larger than the theoretical impact. 
The average temperature of around $40.4\,{}^\circ\mathrm{C}$ before the attack increases to an average temperature of $58.5\,{}^\circ\mathrm{C}$ for Heater 1 at the end of the attack (see upper plot of Figure~\ref{fig:StageIIIforLQGChi2}), which is approximately $1.7\,{}^\circ\mathrm{C}$ degrees larger than the theoretical impact.
Our estimate of the worst-case impact is again very close to the true impact though.
Furthermore, the lower plot of Figure~\ref{fig:StageIIIforLQGChi2} shows us that the attack is still stealthy.
However, as with the MEWMA detector, a visual examination of the detector output would still raise the suspicion of a system operator.

This leads us to conclude that the detector choice can also be a good defense mechanism, since for the LQG controller case we see that the average temperature under the $\chi^2$ detector is approximately $5$ times larger than when the MEWMA detector is used, where both detector thresholds are tuned to the same mean time between false alarms.

\subsection{Noise injection}
\begin{figure}
\centering
\includegraphics[width=0.5\textwidth]{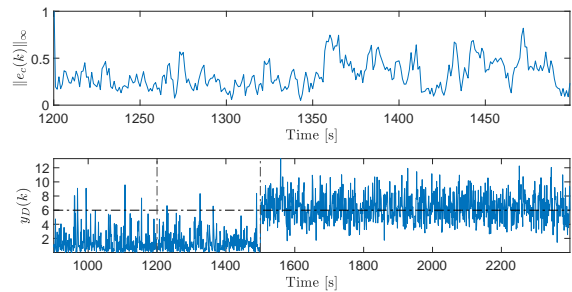}
\caption{The upper plot shows the maximum absolute estimation error of the controller state during Stage I and due to the noise injection this error does not converge to zero. The lower plot shows the detector output without the noise injection until $1199\,\mathrm{s}$ and with noise injection during Stage I and Stage III during the attack. The injected noise does not significantly change the detector output compared to nominal conditions and leads to an immediate detection of the otherwise stealthy sensor attack.}
\label{fig:StageIAndYdLQGChi2WithNoise}
\end{figure}

Finally, we want to discuss noise injection to prevent the attacker from successfully executing Stage I of the attack when a controller is used that has no poles outside of the unit circle.
The basic idea, as proposed in \citep{UmsonstAutomatica21}, is to add an additional noise signal $\nu(k)\sim\mathcal{N}(0,\Sigma_{\nu})$ to the controller state dynamics, i.e.,
\begin{align*}
x_c(k+1)&=A_cx_c(k)+B_cy(k)+\nu(k),\\
u(k)&=C_cx_c(k).
\end{align*}
Since the attacker does not have any access to the noise signal $\nu(k)$, the noise prevents the attacker from obtaining a perfect estimate of the controller state.
Therefore, the attacker will not be able to remain stealthy during Stage II and Stage III of the attack similar to the LQI controller case discussed previously.
A disadvantage of injecting additional noise into the closed-loop is the performance degradation.
The performance degradation will affect the steady-state behavior and controller output by making them noisier, which might damage certain actuators if the signal becomes too noisy.
By changing the normalization matrix of the residual from $\Sigma_r^{-\frac{1}{2}}$ to $(\Sigma_r+CT_c\Sigma_{\nu}C^TT_c^T)^{-\frac{1}{2}}$, the additional noise does not affect the detector performance under nominal conditions.
Hence, in the case of Gaussian process and measurement noise, the mean time between false alarms will not change with the additional noise and this adjusted normalization.
In line with Assumption~\ref{assum:AttackerKnowledge}, we assume the following.
\begin{assumption}
The attacker knows the distribution of $\nu(k)$.
\end{assumption}
Since the LQI controller has an inherent protection against the estimation of the controller state, we look at the LQG controller case when a $\chi^2$ detector is used. 
The experiment has the following time line.
From $900\,\mathrm{s}$ to $1199\,\mathrm{s}$ the noise injection is not used and from $1200\,\mathrm{s}$ on, we start the noise injection with $\Sigma_{\nu}=0.01I_4$.
This is to investigate how the injected noise influences the nominal behavior of the closed-loop system.
Furthermore, the attacker executes Stage I in the interval $[1200,1499]\,\mathrm{s}$, i.e., $k_{\mathrm{I}}=1200$ and $N_{\mathrm{I}}=300$, to demonstrate how the noise injection prevents the controller state estimation.
Then from $1500\,\mathrm{s}$ on the attacker injects the stealthy attack signal of Stage III for $N_{\mathrm{III}}=900\,\mathrm{s}$.
When solving \eqref{eq:StageIIIWorstCaseImpact} with the new normalization for the residual, the estimated worst-case stealthy attack impact is $18.2846$, which is larger than in the case without the noise injection.
However, recall that the injected noise should lead to an early detection of the attack.
\begin{figure}
\centering
\includegraphics[width=0.5\textwidth]{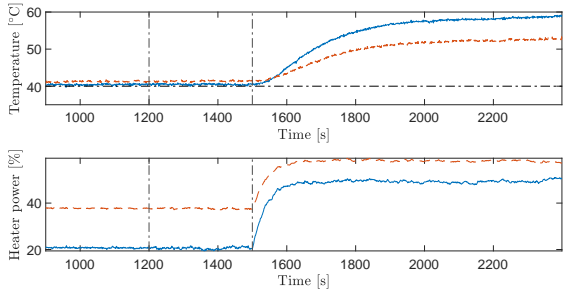}
\caption{The upper plot shows the temperature measurements of the TCLab before the noise injection and during the attack. The lower plot shows the heater power applied to the TCLab before the noise injection and during the attack. We observe that the noise injection does not influence the nominal behavior significantly, while the attack has a similar impact than to the case without noise injection.}
\label{fig:TempAndPowerForLQGChi2WithNoise}
\end{figure}
Figure~\ref{fig:TempAndPowerForLQGChi2WithNoise} shows the measured temperature (upper plot) and the heater power (lower plot) of the noise injection simulation.
The first vertical dash-dotted line from the left marks the start of Stage I and the noise injection, while the second line marks the beginning of Stage III.
We start by noting that the heater power becomes slightly more noisy once the noise injection begins, while the temperature measurements are not much more noisy. 
The overall increase in the noise level in the nominal case is very low, so we conclude that the noise injection does not degrade the system performance much.
In Stage III of the attack, the average temperature of Heater~1 increases from an average of around $40.4\,{}^\circ\mathrm{C}$ to an average of around $58.2\,{}^\circ\mathrm{C}$, which is a very similar increase to the LQG controller case without noise injection under a $\chi^2$ detector.
Now that we looked at the effect of the noise injection on the nominal behavior and the attack impact, let us investigate the effect on Stage I and the stealthiness in Stage III of the attack.
The upper plot of Figure~\ref{fig:StageIAndYdLQGChi2WithNoise} shows the maximum absolute controller state estimation error $\|e_c(k)\|_{\infty}$ during Stage I. Other than in the case without noise injection (upper plot Figure~\ref{fig:StageIforLQGandLQI}) the error does not exponentially converge to zero and there are no signs of a decreasing trend in the error.
The error $e_c(k)$ is, however, unbiased, which means that on average the estimate of the controller state is correct.
If the attacker now launches Stage III of the attack assuming it has a perfect estimate, the detector output will almost immediately cross the threshold as shown in the lower plot of Figure~\ref{fig:StageIAndYdLQGChi2WithNoise}. 
The average value of the detector output during Stage III is $6.3433$, which is larger than the threshold $J_D=5.9915$.
Therefore, we see that the noise injection will lead to a detection of the stealthy attack in Stage III.
Furthermore, we point out that before the attacker injects an additional signal the detector output before and after the noise injection does not look visually different.
Hence, we can conclude that the injection of additional noise to the controller dynamics leads to a detection of the worst-case stealthy attack, while not degrading the performance of the closed-loop system and the detector significantly.

\begin{remark}
In \citep{UmsonstAutomatica21} we propose a convex semi-definite program to obtain an optimal noise covariance matrix $\Sigma_{\nu}$ that takes the performance degradation into account.
However, this optimal noise distribution is under the assumption of a linear system, Gaussian noise, and exact knowledge about the process and measurement noise covariance matrices.
Since these assumptions do not hold in the experiment, we only demonstrated the effectiveness of the noise injection in revealing stealthy attacks in this section without taking the optimality of $\Sigma_{\nu}$ into account.
\end{remark}
\section{Conclusion}
\label{sec:Conclusion}
In this work, we investigate a sensor attack on a feedback system in an experimental setup.
The attack consists of three stages, where the first two stages are a preparation for the third stage by estimating internal loop signals such as the controller and the detector state.
In the third stage, the attacker launches its worst-case attack that remains stealthy to the detector and achieves the largest impact on the plant.
With the experiment, we evaluate if the theoretical results of each stage hold when using real data.
We observe that each of the stages can be successfully completed when a controller with stable dynamics is used, such as the LQG controller.
Furthermore, the theoretically estimated worst-case impacts are close to the actual worst-case impact obtained from the experiment.
By simply adding an integral part to the controller, which is common in practice, the attacker is not able to complete the first stage of the attack exponentially fast. 
Further, without completing the first stage, the attack is detected in the following stages as well in the case of a controller with integral action.
However, the attack impact increases when an integral action is used such as in the LQI controller, which means that if estimating the controller state can be achieved through other means than the ones investigated here, using an integral action can degrade the system performance under the attack.
In case integral action is not a sufficient defense mechanism, we evaluate the noise injection into the controller dynamics as a defense mechanism. We showed that noise injection prevents the attacker from completing the first stage as well, while only slightly degrading the system performance.
In addition to that, we also investigate how the attack impact depends on the detector used and show that the stealthy worst-case impact is smaller when a detector with an internal state is used.

For future research directions, we want to look into different attacker objectives and investigate if detector dynamics can increase the worst-case impact for certain objectives.
Furthermore, the second stage is only examined for detectors with linear dynamics such that we want to investigate if there are detector dynamics, which prevent the attacker from estimating the state of the detector.
Finally, we would like to extend the experimental setup to a more sophisticated testbed for cyber-physical security, where, for example, the sensor and actuator measurements are transmitted wirelessly.
In this testbed more aspects of cyber-physical security could be tested than only the physical impact on the plant.
\section*{Acknowledgments}
This work is supported in part by the Swedish Research Council (grant 2016-00861) and the Swedish Civil Contingencies Agency (grant MSB 2020-09672).

\appendix
\section{Appendix}
\label{appendix:TCLab}
This appendix includes a more detailed exposition of the modeling and the identification of the Temperature Control Lab as well as the controller design.
\subsection{Temperature Control Lab}
\begin{figure}[ht]
\centering
\includegraphics[scale=0.35]{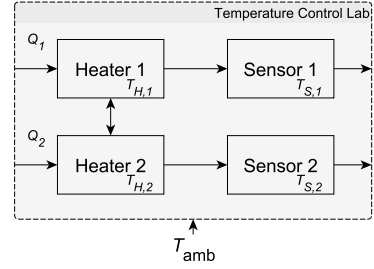}
\caption{Schematic block diagram of the Temperature Control Lab}
\label{figA6:TCLab}
\end{figure}

The Temperature Control Lab (TCLab) is an Arduino-based process with two heaters and two sensors to measure the heater temperatures (see Figure~\ref{figA6:TCLab}). 
The heaters are close to each other such that there exists a coupling between the heater temperatures.
Further, we are able to set the power outputs $Q_1$ for Heater 1 and $Q_2$ for Heater 2 to a value between $0\,\%$ and $100\,\%$.

Several models to describe the TCLab dynamics are investigated in \cite{TCLabModelBenchmark} and the model that is deemed the most accurate is the following physics-based model,
\begin{equation}
\label{eqA6:PhysicsBasedTCLabDynamics}
\begin{aligned}
mc_p\dot{T}_{H,1}(t)&=UA(T_{\text{amb}}-T_{H,1}(t))+\epsilon k_B A (T_{\text{amb}}^4-T_{H,1}(t)^4)\\ &\quad+U_sA_s(T_{H,2}(t)-T_{H,1}(t))\\
&\quad +\epsilon k_B A (T_{H,2}(t)^4-T_{H,1}(t)^4)+\alpha_1Q_1(t)\\
mc_p\dot{T}_{H,2}(t)&=UA(T_{\text{amb}}-T_{H,2}(t))+\epsilon k_B A (T_{\text{amb}}^4-T_{H,2}(t)^4)\\ &\quad-U_sA_s(T_{H,2}(t)-T_{H,1}(t))\\
&\quad -\epsilon k_B A (T_{H,2}(t)^4-T_{H,1}(t)^4)+\alpha_2Q_2(t)\\
\tau_{c,1}\dot{T}_{S,1}(t)&=T_{H,1}(t)-T_{S,1}(t)\\
\tau_{c,2}\dot{T}_{S,2}(t)&=T_{H,2}(t)-T_{S,2}(t).
\end{aligned}
\end{equation}
Here, $T_{H,1}(t)$, $T_{H,2}(t)$, and $T_{\text{amb}}$ are the temperature of Heater 1, of Heater 2, and the ambient temperature, respectively. Further, $T_{S,1}(t)$ and $T_{S,2}(t)$ are the sensor measurements of Heater 1 and Heater 2, respectively.
The temperature unit in \eqref{eqA6:PhysicsBasedTCLabDynamics} is Kelvin, while in the subsequent plots we use degree Celsius to display the temperature measurement.
Table~\ref{tabA6:ModelParam} shows the parameters and their (range of) values of the physical model.
Note that for some parameters in Table~\ref{tabA6:ModelParam} the value is specified by an interval. 
This means these parameters need to be estimated, which we will do next.
\begin{table}
\centering
\caption{Parameters for the physical model (Table 1 in \citet{TCLabModelBenchmark})}
\label{tabA6:ModelParam}
\resizebox{0.5\textwidth}{!}{
\begin{tabular}{|c|c|}
\hline 
Quantity & Value \\ 
\hline 
Heater 1 output ($Q_1$) & $0\,\mathrm{W}$ to $1\,\mathrm{W}$ \\ 
\hline 
Heater 1 factor ($\alpha_1$) & $[0.005,0.02]\,\mathrm{\frac{W}{\%}}$ \\ 
\hline
Heater 2 output ($Q_1$) & $0\,\mathrm{W}$ to $0.73\,\mathrm{W}$ \\ 
\hline 
Heater 2 factor ($\alpha_2$) & $[0.002,0.015]\,\mathrm{\frac{W}{\%}}$ \\ 
\hline 
Heat capacity ($c_p$) & $500\,\mathrm{\frac{J}{kgK}}$ \\ 
\hline 
Surface area not between heaters ($A$) & $1\cdot10^{-3}\,\mathrm{m}^2$ \\ 
\hline 
Surface area not between heaters ($A_s$) & $2\cdot10^{-4}\,\mathrm{m}^2$  \\ 
\hline 
Mass ($m$) & $0.004\,\mathrm{kg}$ \\ 
\hline 
Heat transfer coefficient ($U$) & $[2,30]\,\mathrm{\frac{W}{m^2K}}$ \\ 
\hline 
Heat transfer coefficient between heaters ($U_s$) & $[2,30]\,\mathrm{\frac{W}{m^2K}}$  \\ 
\hline 
Emissivity ($\epsilon$) & 0.9 \\ 
\hline 
Stefan Boltzmann constant $(k_B)$ & $5.67\cdot10^{-8}\,\mathrm{\frac{W}{m^2K^2}}$ \\ 
\hline 
Conduction time constants ($\tau_{c,1},\tau_{c,2})$ & $[15,40]\,\mathrm{s}$ \\ 
\hline 
\end{tabular} 
}
\end{table}

\begin{figure}[ht]
\centering
\includegraphics[width=0.45\textwidth]{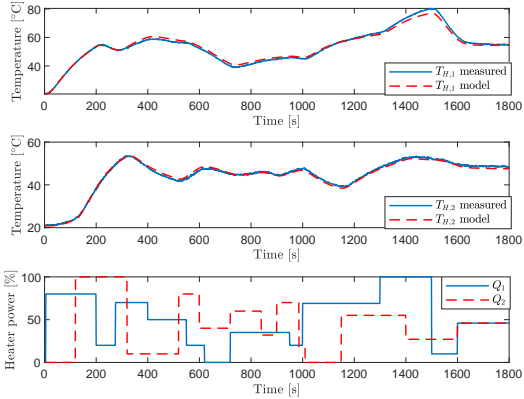}
\caption{Trajectories used for the estimation of unknown parameters}
\label{figA6:IdentificationData}
\end{figure}

\begin{figure}[ht]
\centering
\includegraphics[width=0.45\textwidth]{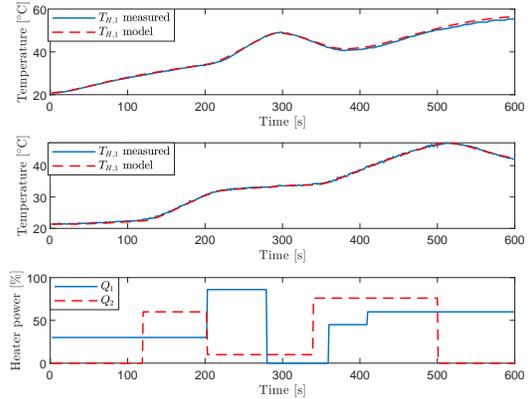}
\caption{Trajectories used for the validation of our estimated model parameters}
\label{figA6:ValidationData}
\end{figure}

The code to interact with the TCLab and for parameter estimation can be found at \cite{TCLabResources}. 
The code from \cite{TCLabResources} is adjusted to estimate both $\tau_{c,1}$ and $\tau_{c,2}$ instead of assuming that $\tau_{c,1}=\tau_{c,2}$ as in \cite{TCLabModelBenchmark}. 
To estimate the unknown parameters, we apply piecewise constant control inputs to the TCLab and use the measured output signals for parameter estimation (see Figure~\ref{figA6:IdentificationData}), where we set the ambient temperature, $T_{\mathrm{amb}}$, to be the average of the first measurement of the two temperature sensors.
With the control inputs and the sensor measurements we want to determine $\alpha_1$, $\alpha_2$, $U$, $U_s$, $\tau_{c,1}$, and $\tau_{c,2}$ by solving the following optimization problem as in the code provided in \cite{TCLabResources},
\begin{equation}
\label{eqA6:ObjectiveForParameterEstimation}
\begin{aligned}
\min_{\substack{\alpha_1,\alpha_2,U,\\U_s,\tau_{c,1},\tau_{c,2}}}&\sum_{i=1}^{N_{\mathrm{id}}} \left(\left(\frac{T_{S,1\text{meas}}(t_i)-T_{S,1}(t_i)}{T_{S,1\text{meas}}(t_i)}\right)^2\right.\\
&\left.\quad \ \, +\left(\frac{T_{S,2\text{meas}}(t_i)-T_{S,2}(t_i)}{T_{S,2\text{meas}}(t_i)}\right)^2\right),
\end{aligned}
\end{equation}
where $t_i$ are the time points at which we measured the temperature, $N_{\mathrm{id}}$ is the number of measurements, and $T_{S,1\text{meas}}(t_{i})$ and $T_{S,2\text{meas}}(t_{i})$ are the measured temperatures for the sensors of Heater 1 and Heater 2, respectively.
To estimate the unknown parameters, we collected data for thirty minutes, i.e., $N_{id}=1800$, and used a sampling time of one second, such that $t_{i}-t_{i-1}=1\,\mathrm{s}$ (see Figure~\ref{figA6:IdentificationData}).
With the data in Figure~\ref{figA6:IdentificationData}, the parameters that minimizes the objective in \eqref{eqA6:ObjectiveForParameterEstimation} are $\alpha_1=0.00854\,\mathrm{\frac{W}{\%}}$, $\alpha_2=0.00480\,\mathrm{\frac{W}{\%}}$, $U=4.05\,\mathrm{\frac{W}{m^2K}}$, $U_s=26.44\,\mathrm{\frac{W}{m^2K}}$, $\tau_{c,1}=25.16\,\mathrm{s}$, and $\tau_{c,2}=22.50\,\mathrm{s}$. The value of the objective function with the data from Figure~\ref{figA6:IdentificationData} is $1.46$.
Finally, to validate our estimated model parameters, we recorded ten more minutes, i.e., $N_{\mathrm{id}}=600$, of data and compared it with our the output of the theoretical model with the estimated parameters (see Figure~\ref{figA6:ValidationData}).
We observe that our theoretical model tracks the validation data well and the objective function in \eqref{eqA6:ObjectiveForParameterEstimation} has a value of $0.13653$ with the data from Figure~\ref{figA6:ValidationData} and the previously estimated parameters.

\subsection{Controller design}
Now that we introduced the TCLab, its dynamics, and identified the parameters, we want to control the TCLab around a certain temperature.
For that we design a linear controller, which utilizes a linear discrete-time model of the TCLab.
We would like to point out that we want to design a controller that performs satisfactorily around the steady state and we do not consider more constraints on the rise time and the settling time, for example.
 
To find a linear model, we linearize the non-linear physics-based dynamics \eqref{eqA6:PhysicsBasedTCLabDynamics} around a steady-state value.
We observe that $\lim_{t\rightarrow\infty}T_{S,i}(t)=\lim_{t\rightarrow\infty}T_{H,i}(t)=T_{H,i\infty}$ for $i\in\lbrace 1,2\rbrace$, thus we only need to find the steady-state values of the heater temperatures and the heater output, denoted as $T_{H,1\infty}$, $T_{H,2\infty}$, $Q_{1\infty}$, and $Q_{2\infty}$, respectively.
The equations to find the steady-state values are given by
\begin{align*}
0&=UA(T_{\text{amb}}-T_{H,1\infty})+\epsilon k_B A (T_{\text{amb}}^4-T_{H,1\infty}^4)\\
&\quad +U_sA_s(T_{H,2\infty}-T_{H,1\infty})+\epsilon k_B A (T_{H,2\infty}^4-T_{H,1\infty}^4)\\
&\quad +\alpha_1Q_{1\infty},\\
0&=UA(T_{\text{amb}}-T_{H,2\infty})+\epsilon k_B A (T_{\text{amb}}^4-T_{H,2\infty}^4)\\
&\quad -U_sA_s(T_{H,2\infty}-T_{H,1\infty})-\epsilon k_B A (T_{H,2\infty}^4-T_{H,1\infty}^4)\\
&\quad+\alpha_2Q_{2\infty},
\end{align*}
where we assume that the ambient temperature $T_{\text{amb}}$ is known and fixed.
Alternatively, we can also add the two steady-state equation and also subtract them from each other to obtain the subsequent steady state equations,
\begin{align*}
0&=UA(2T_{\text{amb}}-T_{H,1\infty}-T_{H,2\infty})\\
&\quad +\epsilon k_B A (2T_{\text{amb}}^4-T_{H,1\infty}^4-T_{H,2\infty}^4)+\alpha_1Q_{1\infty}+\alpha_2Q_{2\infty},\\
0&=3UA(T_{H,2\infty}-T_{H,1\infty})+3\epsilon k_B A (T_{H,2\infty}^4-T_{H,1\infty}^4)\\
&\quad +\alpha_1Q_{1,\infty}-\alpha_2Q_{2\infty}.
\end{align*}
Now, we can fix the desired steady-state temperature $T_{H1,\infty}$ and $T_{H2,\infty}$, and solve for $Q_{1\infty}$ and $Q_{2\infty}$. 
With $T_{H1,\infty}=T_{H2,\infty}=T_{H,\infty}$ we derive that $Q_{2\infty}=\frac{\alpha_1}{\alpha_2}Q_{1\infty}$ and 
\begin{align*}
Q_{1\infty}&=\frac{1}{\alpha_1}\left( UA(T_{H,\infty}-T_{\text{amb}})+\epsilon k_B A (T_{H,\infty}^4-T_{\text{amb}}^4)\right),
\end{align*}
are the necessary steady-state inputs to achieve $T_{H1,\infty}=T_{H2,\infty}=T_{H,\infty}$.
Note that we need $T_{H,\infty}\geq T_{\mathrm{amb}}$ to ensure that both $Q_{1\infty}\geq 0$ and $Q_{2\infty}\geq 0$.

\begin{figure}[ht]
\centering
\includegraphics[width=0.45\textwidth]{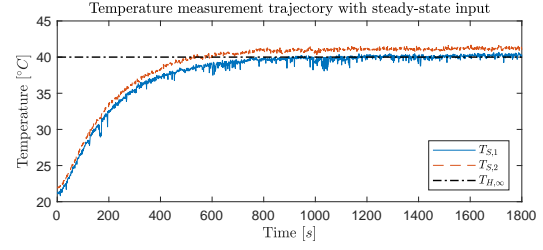}
\caption{Application of the steady-state inputs to achieve $T_{H,1\infty}=T_{H,2\infty}=40\,{}^\circ \mathrm{C}$}
\label{figA6:SteadyStateExperimentNewModel}
\end{figure}

Figure~\ref{figA6:SteadyStateExperimentNewModel} shows a steady-state input that theoretically results in the temperatures being the same value, where the ambient temperature is $T_{\text{amb}}=21\,{}^\circ \mathrm{C}$.
In this figure, we use $T_{H,\infty}=40\,{}^\circ \mathrm{C}$ such that the required steady-state inputs are $Q_{1\infty}=21.73\,\%$ and $Q_{2\infty}=38.72\,\%$.
For $k\geq 1000$ the average temperature for the first heater after the transient is approximate $40\,{}^\circ \mathrm{C}$, while the second heater reaches $41.12\,{}^\circ \mathrm{C}$.
Hence, we observe that Heater 1 reaches the desired steady-state temperature of $40\,{}^{\circ} \mathrm{C}$, while the second heater is around $1.2\,{}^\circ\mathrm{C}$ warmer than $40\,{}^{\circ} \mathrm{C}$.
We, further, note that the heaters do not reach the same temperature.
Despite performing well with the validation data as shown in Figure~\ref{figA6:ValidationData}, we observe here that our estimated model parameters do not seem to be able to model the steady-state dynamics well.
Hence, our theoretical model does not describe the TCLab perfectly. 

Next, we look into the feedback controller design to achieve a better control around the desired steady-state value than the feedforward injection of the steady-state values.

First, we linearize the TCLab equations with the steady-state values as seen in \eqref{eqA:LinearizedTCLabDynamics},
\begin{figure*}
\begin{equation}
\label{eqA:LinearizedTCLabDynamics}
\begin{aligned}
\dot{x}(t)&=\begin{bmatrix}
-\frac{UA}{mc_p}-4\frac{\epsilon k_B A}{mc_p}T_{H,1\infty}^3 & \frac{U_sA_s}{mc_p}+4\frac{\epsilon k_B A}{mc_p}T_{H,2\infty}^3 & 0 & 0 \\
 \frac{U_sA_s}{mc_p}+4\frac{\epsilon k_B A}{mc_p}T_{H,1\infty}^3  & -\frac{UA}{mc_p}-4\frac{\epsilon k_B A}{mc_p}T_{H,2\infty}^3 & 0 & 0 \\ \frac{1}{\tau_{c,1}} & 0 & -\frac{1}{\tau_{c,1}} & 0\\
0 & \frac{1}{\tau_{c,2}} & 0 & -\frac{1}{\tau_{c,2}}
\end{bmatrix}x(t)
+\begin{bmatrix}
\alpha_1 & 0\\
0 & \alpha_2\\
0 & 0\\
0 & 0
\end{bmatrix}u(t),\\
y(t)&=\begin{bmatrix}
0 & 0 & 1 & 0\\
0 & 0 & 0 & 1
\end{bmatrix}x(t),
\end{aligned}
\end{equation}
\end{figure*}
where $x(t)=[T_{H,1}-T_{H,1\infty}, T_{H,2}-T_{H,2\infty}, T_{S,1}-T_{H,1\infty}, T_{S,2}-T_{H,2\infty}]^\top $, $u(t)=[Q_1-Q_{1,\infty}, Q_2-Q_{2,\infty}]^\top $, and we did not linearize around the ambient temperature since it is assumed to be constant. 
Note that with the assumption of a constant ambient temperature, the linearized model does not depend on the ambient temperature itself.

\begin{figure}[ht]
\centering
\includegraphics[scale=0.35]{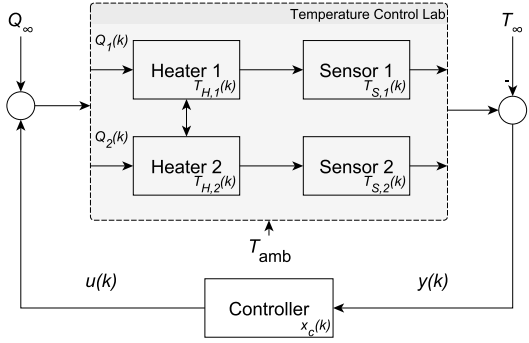}
\caption{The TCLab process equipped with a linear controller that controls the TCLab around the steady state, where $Q_{\infty}=[Q_{1,\infty}, Q_{2,\infty}]^\top $ and $T_{\infty}=[T_{H,1\infty},T_{H,2\infty}]^\top $.}
\label{figA6:TCLabWithController}
\end{figure}

Wanting to reach $T_{H1,\infty}=T_{H2,\infty}=40\,{}^\circ \mathrm{C}$, and discretizing the linearized equations with a sampling time of $T_s=1\,\mathrm{s}$ leads to our linearized discrete-time model,
\begin{align*}
x(k+1)&=Ax(k)+Bu(k),\\
y(k)&=Cx(k),
\end{align*}
where
\begin{align*}
A&=\begin{bmatrix} 
	0.9784  &  0.0113  &       0  &       0\\
    0.0113  &  0.9784  &       0  &       0\\
    0.0385  &  0.0002  &  0.9610  &       0\\
    0.0002  &  0.0430  &       0  &  0.9565\end{bmatrix},\\
B&=\begin{bmatrix} 
    0.0085  &  0\\
    0		&  0.0047\\
    0.0002  &  0\\
    0	  	&  0.0001\end{bmatrix},\ \mathrm{and}\ C=\begin{bmatrix}
0 & 0 & 1 & 0\\
0 & 0 & 0 & 1
\end{bmatrix}.
\end{align*}
Next, we design two linear controllers for the TCLab based on the linearized discrete-time model, which should control the TCLab around its steady state.
In the following, we set $T_{\text{amb}}=21\,{}^{\circ}\mathrm{C}$, since the experiments are conducted at room temperature.
The closed-loop block diagram is shown in Figure~\ref{figA6:TCLabWithController}.

The first controller we design is a linear quadratic Gaussian controller (LQG controller), which is a combination of a linear quadratic regulator and a Kalman filter.
The linear quadratic regulator designs a state-feedback control law such that the cost function, $\sum_{k=0}^\infty x(k)^\top Q_x x(k)+u(k)^\top R_u u(k)$ is minimized, where $Q_x\in\mathbb{R}^{4\times 4}$ and $R_u\in\mathbb{R}^{2\times 2}$ are cost matrices that penalize the state and the controller input, respectively. Since we can only obtain a noisy measurement $y(k)$, a steady-state Kalman filter is used to estimate the state $x(k)$. The Kalman filter minimizes the mean square error, $\mathbb{E}\lbrace\|x(k)-\hat{x}(k)\|_2^2\rbrace$, where $\hat{x}(k)$ is the Kalman filter's estimate of the plant's state.
Therefore, the controller input is given by $u(k)=-K\hat{x}(k)$ and the optimal controller gain $K$ is given by
\begin{align}
\label{eqA6:ControllerGainLQR}
K=(R_u+B^\top PB)^{-1}B^\top PA
\end{align}
and $P$ is the solution to the Riccati equation
\begin{equation}
\label{eqA6:RiccatiLQR}
\begin{aligned}
P_K&=A^\top P_KA+Q_x\\
&\quad +A^\top P_KB(R_u+B^\top P_KB)^{-1}B^\top P_KA.
\end{aligned}
\end{equation}
The steady-state Kalman filter dynamics are
\begin{align*}
\hat{x}(k+1)=(A-LC)\hat{x}(k)+Ly(k),
\end{align*}
where $L$ is the steady-state Kalman gain given by
\begin{align*}
L=AP_LC^\top (CP_LC^\top +\Sigma_v)^{-1}
\end{align*}
and $P_L$ is the solution to the Riccati equation
\begin{align*}
P_L=AP_LA^\top +\Sigma_w+AP_LC^\top (CP_LC^\top +\Sigma_v)^{-1}CP_LA^\top .
\end{align*}
Here, $\Sigma_w\in\mathbb{R}^{4\times 4}$ and $\Sigma_v\in\mathbb{R}^{2\times 2}$ are the covariance matrices of the process noise affecting the linear system dynamics and of the measurement noise affecting the measurements.

When designing the LQG controller, the parameters $Q_x$, $R_u$, $\Sigma_w$, and $\Sigma_v$ are our design variables.
We use the guidelines provided by \cite{LQGDesignPhilosophy} to tune these design variables.
Since the Kalman filter estimate is used to determine the control signal, it is advised to not choose the matrices, $Q_x$ and $R_u$, for the LQR design and the noise covariance matrices, $\Sigma_w$ and $\Sigma_v$, for the Kalman filter design independently \cite{LQGDesignPhilosophy}.

Since the theoretical steady state is not reached when applying the steady-state input (see Figure~\ref{figA6:SteadyStateExperimentNewModel}), we want the controller to have a tighter control around the steady state and therefore we choose a larger cost matrix for the state in the LQR problem. 
Since the steady-state control input is relatively small, we also decide to not penalize the control input with a large cost matrix, which leads to the following choice of state and control input cost matrices, $Q_x=10I_{4}$ and $R_u=2I_2$, respectively.
For the Kalman filter, the value of the process noise covariance matrix can be interpreted as how much we trust the theoretical model. 
Hence, a large process noise covariance matrix compared to the measurement noise covariance matrix means that we trust the sensor measurements more than our model. 
Since the theoretical result does not match the real steady-state value in Figure~\ref{figA6:SteadyStateExperimentNewModel}, we set the process noise matrix as $\Sigma_w=5I_4$. 
Further, we know that the sensors have an accuracy of $\pm1\,{}^\circ\mathrm{C}$, so we choose the measurement noise matrix as $\Sigma_v=I_{2}$.

\begin{figure}[ht]
\centering
\includegraphics[width=0.45\textwidth]{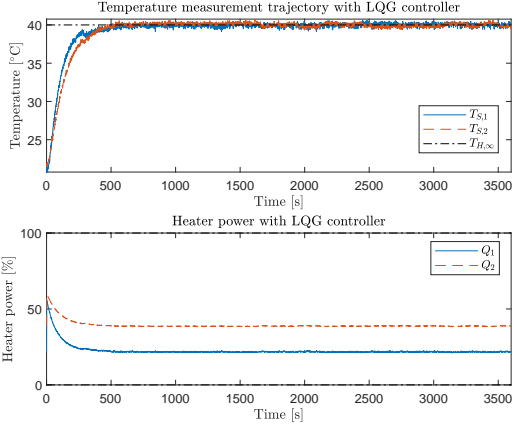}
\caption{Using an LQG controller to control the TCLab temperatures around their steady-state value of $40\,{}^\circ\mathrm{C}$, where the upper plot show the temperature trajectories, while the lower plot shows the heater power.}
\label{figA6:LQGController}
\end{figure}

We test the designed LQG controller in the following. We apply the steady-state input from $t\geq 0\,\mathrm{s}$ on and activate the controller at the same time.
The trajectories of the temperature measurements and the heater power can be seen in Figure~\ref{figA6:LQGController}. 
We observe that the controller leads to a spike in the heater power, which leads to a faster convergence close to the steady-state value than in Figure~\ref{figA6:SteadyStateExperimentNewModel}.
Furthermore, the spike in the control input does not exceed the maximum control input of $100\,\%$.
After reaching their steady state both temperatures remain at an average temperature of approximately $40\,{}^\circ\mathrm{C}$ for $t\in[900,3600]\,\mathrm{s}$, which shows us that the controller successfully manages to keep the temperatures at the desired steady-state temperature.
The mean value of the heater power for $t\in[900,3600]\,\mathrm{s}$ is $21.73\,\%$ for Heater 1 and $38.71\,\%$ for Heater 2, which is close the steady-state inputs $Q_{1,\infty}=21.73\,\%$ and $Q_{2,\infty}=38.71\,\%$.

\begin{figure}[ht]
\centering
\includegraphics[width=0.45\textwidth]{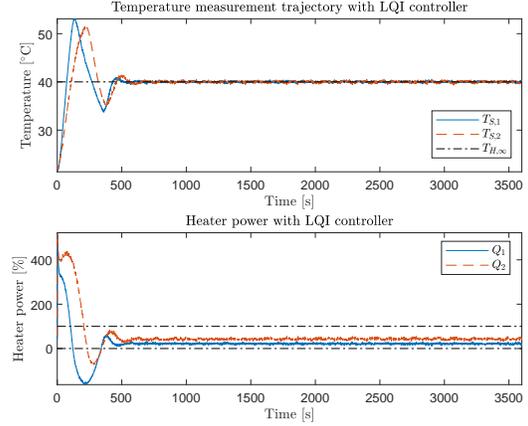}
\caption{Using an LQI controller to control the TCLab temperatures around their steady-state value of $40\,{}^\circ\mathrm{C}$, where the upper plot show the temperature trajectories, while the lower plot shows the controller input.}
\label{figA6:LQIController}
\end{figure}

Next, we add an integral action to the LQG controller. The LQG controller with integral action is subsequently called LQI controller.
The control input of the LQI controller is given by
\begin{align*}
u(k)&=-K_{\mathrm{int}}\begin{bmatrix}
\hat{x}(k)\\
x_{\mathrm{int}}(k)
\end{bmatrix},\\
x_{\mathrm{int}}(k+1)&=x_{\mathrm{int}}(k)+T_s(d(k)-y(k)),
\end{align*}
where $x_{\mathrm{int}}(k)\in\mathbb{R}^{n_y}$ is the state of the integrator and $d(k)$ is the desired reference, which the output $y(k)$ should track.
Since $y(k)$ represents the deviation from the reference steady-state temperature, it should be close to zero such that $d(k)=0$ is used in the following.
To design the controller matrix $K_{\mathrm{int}}$, we use again equations \eqref{eqA6:ControllerGainLQR} and \eqref{eqA6:RiccatiLQR}, but we use $A_{\mathrm{aug}}$ and $B_{\mathrm{aug}}$ instead of $A$ and $B$, where
\begin{align*}
A_{\mathrm{aug}}=\begin{bmatrix}
A & 0\\-T_sC & I
\end{bmatrix}\ 
\mathrm{and}\ 
B_{\mathrm{aug}}=\begin{bmatrix}
B\\0
\end{bmatrix},
\end{align*}
are the system and input matrix of the plant augmented with the integrator state.
Since the augmented system has two more states, we need to change the cost matrix $Q_x$ in \eqref{eqA6:RiccatiLQR} as well.
The new cost matrix for the augmented system state is chosen as $Q_{x,{\mathrm{int}}}=\mathrm{diag}(10I_4,2I_{2})$, while we still use $R_u=2I_{2}$ for the cost matrix of the controller input.
Since we have access to the integrator states, we do not need to re-calculate the observer gain $L$.
Using the LQI controller, we obtain temperature trajectories that reach the steady state (see Figure~\ref{figA6:LQIController}).
Initially the controller leads to an oscillatory behavior until the steady state is reached. 
The reason for these oscillations is the large deviation from the steady-state value at the beginning of the experiment, which integrates up and results in an integrator windup. 
Anti-windup schemes can present a remedy to the oscillatory behavior but are not investigated here.
For more information on anti-windup schemes we refer the reader to \cite{AntiWindupTutorial}.
Furthermore, we see that in the beginning the heater power should have values below $0\,\%$ and above $100\,\%$, but the saturation of the heater power limits the values to the interval between $0\,\%$ and above $100\,\%$.
Once the TCLab reaches temperatures around the desired steady-state value, we see that the control input is inside the interval between $0\,\%$ and above $100\,\%$ and that both heaters have the same temperature. 
The average heater power in the interval $[900,3600]\,\mathrm{s}$ is $21.54\,\%$ and $43.24\,\%$, for Heater 1 and Heater 2, respectively. 
We see that the average heater power during steady state for Heater 2 is larger than the average control input for Heater 2 when using the LQG controller in Figure~\ref{figA6:LQGController}, while the heater power for Heater 1 is similar.
The reason for that is that the LQI controller can adjust to changes in, for example, the ambient temperature due to the integral action. Therefore, the LQI controller is able to keep the temperatures at the desired steady-state value.
\bibliographystyle{plainnat}        
\bibliography{bibExperimentalValidation}           

\end{document}